\documentclass[twocolumn,showpacs,preprintnumbers,amsmath,amssymb,superscriptaddress]{revtex4}

\usepackage{graphicx}
\usepackage{dcolumn}
\usepackage{bm}
\begin{document}
\preprint{APS/123-QED}
\title{Mass Measurements and Implications for the Energy of the High-Spin Isomer in $^{94}$Ag}
\author{A.~Kankainen} \email{anu.kankainen@phys.jyu.fi}  
\affiliation{Department of Physics, University of Jyv\"askyl\"a, P.O. Box 35, FI-40014 University of Jyv\"askyl\"a, Finland}
\author{V.-V.~Elomaa}
\affiliation{Department of Physics, University of Jyv\"askyl\"a, P.O. Box 35, FI-40014 University of Jyv\"askyl\"a, Finland}
\author{L.~Batist}
\affiliation{Petersburg Nuclear Physics Institute, 188300 Gatchina, Russia}
\author{S.~Eliseev}
\affiliation{Petersburg Nuclear Physics Institute, 188300 Gatchina, Russia}
\affiliation{Gesellschaft f\"ur Schwerionenforschung mbH, Planckstra\ss e1, D-64291 Darmstadt, Germany}
\author{T.~Eronen}
\affiliation{Department of Physics, University of Jyv\"askyl\"a, P.O. Box 35, FI-40014 University of Jyv\"askyl\"a, Finland}
\author{U.~Hager} \altaffiliation[Present address:]{TRIUMF 4004 Wesbrook Mall, Vancouver, BC, V6T 2A3 Canada} 
\affiliation{Department of Physics, University of Jyv\"askyl\"a, P.O. Box 35, FI-40014 University of Jyv\"askyl\"a, Finland}
\author{J.~Hakala}
\affiliation{Department of Physics, University of Jyv\"askyl\"a, P.O. Box 35, FI-40014 University of Jyv\"askyl\"a, Finland}
\author{A.~Jokinen}
\affiliation{Department of Physics, University of Jyv\"askyl\"a, P.O. Box 35, FI-40014 University of Jyv\"askyl\"a, Finland}
\author{I.D.~Moore}
\affiliation{Department of Physics, University of Jyv\"askyl\"a, P.O. Box 35, FI-40014 University of Jyv\"askyl\"a, Finland}
\author{Yu.N.~Novikov}
\affiliation{Petersburg Nuclear Physics Institute, 188300 Gatchina, Russia}
\affiliation{Gesellschaft f\"ur Schwerionenforschung mbH, Planckstra\ss e1, D-64291 Darmstadt, Germany}
\author{H.~Penttil\"a}
\affiliation{Department of Physics, University of Jyv\"askyl\"a, P.O. Box 35, FI-40014 University of Jyv\"askyl\"a, Finland}
\author{A.~Popov}
\affiliation{Petersburg Nuclear Physics Institute, 188300 Gatchina, Russia}
\author{S.~Rahaman}
\affiliation{Department of Physics, University of Jyv\"askyl\"a, P.O. Box 35, FI-40014 University of Jyv\"askyl\"a, Finland}
\author{S.~Rinta-Antila}\altaffiliation[Present address:]{Department of Physics, University of Liverpool, Liverpool L69 7ZE, 
UK} 
\affiliation{Department of Physics, University of Jyv\"askyl\"a, P.O. Box 35, FI-40014 University of Jyv\"askyl\"a, Finland}
\author{J.~Rissanen}
\affiliation{Department of Physics, University of Jyv\"askyl\"a, P.O. Box 35, FI-40014 University of Jyv\"askyl\"a, Finland}
\author{A.~Saastamoinen}
\affiliation{Department of Physics, University of Jyv\"askyl\"a, P.O. Box 35, FI-40014 University of Jyv\"askyl\"a, Finland}
\author{D.M.~Seliverstov}
\affiliation{Petersburg Nuclear Physics Institute, 188300 Gatchina, Russia}
\author{T.~Sonoda} \altaffiliation[Present address:]{Atomic Physics Laboratory, RIKEN, 2-1 Hirosawa, Wako, Saitama 351-0198, 
Japan}
\affiliation{Department of Physics, University of Jyv\"askyl\"a, P.O. Box 35, FI-40014 University of Jyv\"askyl\"a, Finland}
\author{G.~Vorobjev}
\affiliation{Petersburg Nuclear Physics Institute, 188300 Gatchina, Russia}
\affiliation{Gesellschaft f\"ur Schwerionenforschung mbH, Planckstra\ss e1, D-64291 Darmstadt, Germany}
\author{C.~Weber}
\affiliation{Department of Physics, University of Jyv\"askyl\"a, P.O. Box 35, FI-40014 University of Jyv\"askyl\"a, Finland}
\author{J.~\"Ayst\"o}
\affiliation{Department of Physics, University of Jyv\"askyl\"a, P.O. Box 35, FI-40014 University of Jyv\"askyl\"a, Finland}

\date{\today}

\begin{abstract}
Nuclides in the vicinity of $^{94}$Ag have been studied with the Penning trap mass spectrometer JYFLTRAP at the Ion-Guide 
Isotope Separator On-Line. The masses of the two-proton-decay daughter $^{92}$Rh and the beta-decay daughter $^{94}$Pd of the 
high-spin isomer in $^{94}$Ag have been measured, and the masses of $^{93}$Pd and $^{94}$Ag  have been deduced. When combined 
with the data from the one-proton- or two-proton-decay experiments, the results lead to contradictory mass excess values for 
the high-spin isomer in $^{94}$Ag, $-46370(170)$ or $-44970(100)~\text{keV}$, corresponding to excitation energies of 
$6960(400)$ or $8360(370)~\text{keV}$, respectively.
\end{abstract}


\maketitle

Exotic decay modes of the highest-spin isomer in the $N~=~Z$ nuclide $^{94}$Ag have been puzzling nuclear physicists for 
several years. The properties of this isomer are unprecedented in the entire known Segr\'e chart and have resulted in a series 
of intensive studies \cite{Sch94, La02, Ple04, Muk04, Muk05, Muk06, Pec07}. Because of its high excitation energy, this isomer 
with a half-life of $T_{1/2}=0.39(4)~\text{s}$ \cite{Muk04} and spin ($21^+$), can decay via $\beta$ decay or $\beta$-delayed 
proton emission up to three protons or directly via one- or two-proton decay. Recently, the possibility of the two-proton 
decay mode was questioned \cite{Pec07}. In order to uncover the nature of this isomer and its possible decay modes, the decay 
energies, i.e., the masses of the nuclei involved, should be experimentally determined, since they are based on extrapolations 
of systematic trends in the Atomic Mass Evaluation 2003 (AME) \cite{Aud03}. In this Letter, we present the results from 
precision mass measurements around $^{94}$Ag and discuss their impact on the high-spin isomer in $^{94}$Ag.  

The masses of $^{84}$Y, $^{87}$Zr, $^{88,89}$Mo, $^{88-92}$Tc, $^{90-94}$Ru, $^{92-95}$Rh, and $^{94-96}$Pd have been measured 
with JYFLTRAP \cite{Kol04}, a double Penning trap at the IGISOL (Ion-Guide Isotope Separator On-Line) \cite{Ays01}, in a joint 
project with SHIPTRAP ion trap facility at GSI \cite{SHIP1, SHIP2} to investigate nuclides in the region of the $rp$- and 
possible $\nu p$-process paths. The results of the project will be published in a separate paper \cite{Elo08}. Of the measured 
nuclides, the masses of the two-proton-decay daughter $^{92}$Rh and the beta-decay daughter $^{94}$Pd given in 
Table~\ref{tab:1} are essential for the study of $^{94}$Ag. These nuclides were produced via heavy-ion fusion-evaporation 
reactions induced by a $^{40}$Ca beam on a $^{\text{nat}}$Ni target at the IGISOL where almost all reaction products end up at 
a $q=1^+$ charge state. The ions were accelerated to 30 keV, mass-separated and delivered to a radio frequency quadrupole 
(RFQ) ion beam cooler and buncher \cite{Nie01}. The RFQ transferred the ion bunches to JYFLTRAP which consists of two Penning 
traps situated in a homogeneous magnetic field. The first trap is used for isobaric purification by mass-selective buffer-gas 
cooling \cite{Sav91} and the second trap is dedicated to high-precision mass measurements via a time-of-flight cyclotron 
frequency determination \cite{Gra80}. By measuring the time of flight as a function of excitation frequency, the cyclotron 
frequency ${\nu _c = qB/(2\pi m)}$ can be determined. The magnetic field $B$ is calibrated by measuring well-known reference 
masses, for example, the nuclides $^{85}$Rb and $^{94}$Mo in this work. From the measured cyclotron frequency ratios 
$\nu_{c,ref}/\nu_c$, the mass excess values $\Delta$ were derived.  With this method, a typical mass uncertainty below 10 keV 
was achieved for the nuclides. 

\begin{table*}[!]
\caption{\label{tab:1} Mass excess values $\Delta$ for $^{92}$Rh and $^{94}$Pd. The mass excess values were derived using the 
mass of a reference nuclide as given in Ref.~\cite{Aud03}. In column four,``\#" indicates a value that is derived from 
experimental, systematic trends \cite{Aud03}.}
\begin{ruledtabular}       
\begin{tabular}{ccccc}
Nuclide & Ref. atom & $\Delta_\text{JYFL}$ (keV) &  $\Delta_\text{AME}$ (keV) & JYFL$-$AME (keV)\\
\hline
$^{92}$Rh & $^{85}$Rb & $-$62998.6(4.3)\footnotemark[1]& $-$63360(400)\# & 361(400)\\
$^{94}$Pd & $^{94}$Mo & $-$66097.9(4.7)& $-$66350(400)\# & 252(400)\\
\end{tabular}
\end{ruledtabular}
\footnotetext[1]{An average of the JYFLTRAP and SHIPTRAP values.}
\end{table*}

\begin{figure}[!]
\center{
\resizebox{0.4\textwidth}{!}{
\includegraphics{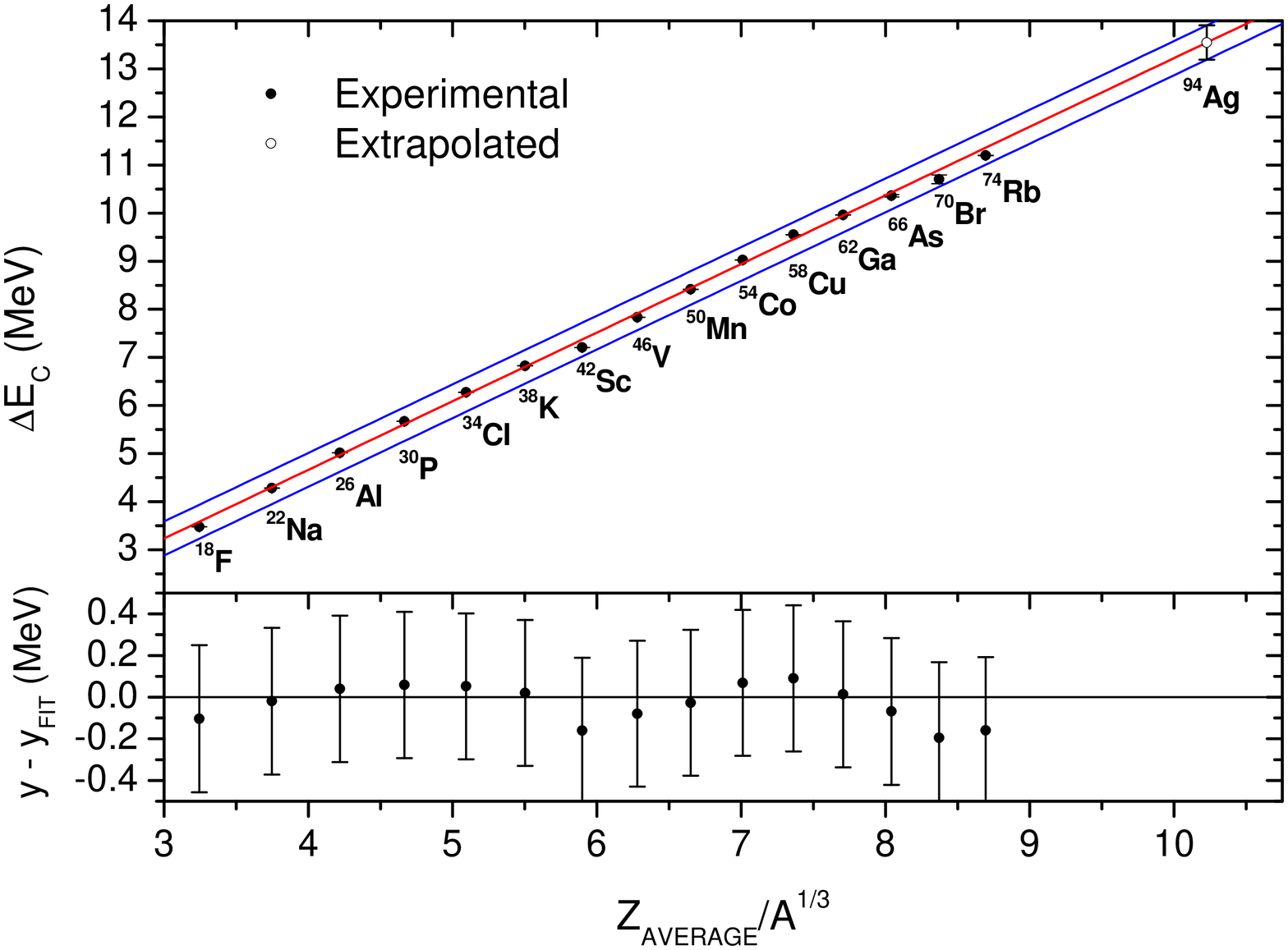}}
\caption{(color online). Experimental Coulomb displacement energies for odd-odd ${N=Z}$ nuclei. A 68~\% prediction band gives 
$\Delta E_\text{C}=13550(360)~\text{keV}$ for $^{94}$Ag. The $Q_\text{EC}$ values of $^{26}$Al \cite{Ero06a}, $^{42}$Sc 
\cite{Ero06a}, $^{46}$V \cite{Ero06a}, $^{50}$Mn \cite{Ero08}, $^{54}$Co \cite{Ero08}, $^{62}$Ga \cite{Ero06b}, and $^{74}$Rb 
\cite{Kel04} as well as the masses of $^{22}$Na \cite{Muk08}, $^{38}$K \cite{Yaz07} and $^{66}$As \cite{Sch07} are from recent 
Penning trap measurements. The value for $^{34}$Cl is taken from the compilation of Hardy and Towner \cite{Har05} and for 
$^{70}$Br from Refs.~\cite{Har05, Kar04}. All other values are from Ref.~\cite{Aud03}. The residuals of the fit together with 
the errors representing the 68~\% prediction band are shown in the lower panel.}
\label{fig:coulomb}       
}
\end{figure}

Although the mass of $^{94}$Ag was not directly measured, the mass of its $\beta$-decay daughter, $^{94}$Pd, was determined in 
this work. Since the ground state of $^{94}$Ag is presumably a $T=1$ isobaric analog state \cite{Jan05}, the $Q_\text{EC}$ 
value of $^{94}$Ag can be estimated quite accurately from the Coulomb displacement energy $\Delta 
E_\text{C}=Q_\text{EC}+\Delta_\text{nH}$, where $\Delta_\text{nH}$ is the neutron-hydrogen mass difference \cite{Aud03}. Thus, 
the mass of $^{94}$Ag can be obtained from the $Q_\text{EC}$ value and the mass of $^{94}$Pd. Experimental Coulomb 
displacement energies have been investigated as a function of $Z_\text{{average}}/A^{1/3}$ in Ref.~\cite{Ant97}, where 
$Z_\text{average}=\frac{1}{2}(Z_\text{mother}+Z_\text{daughter})$. However, the linear fit results in a significant deviation 
for the heaviest known $T=1$ nucleus $^{74}$Rb. Therefore, a new fit to the Coulomb displacement energies based on the current 
experimental $Q_\text{EC}$ values of the odd-odd ${N~=~Z}$ nuclei is shown in Fig.~\ref{fig:coulomb}. This new fit includes 
nuclides such as $^{62}$Ga \cite{Ero06b}, $^{66}$As \cite{Sch07} and $^{74}$Rb \cite{Kel04} for which precise Penning trap 
measurements are now available. With a 68~\% prediction band, the fit gives $\Delta E_\text{C}=13550(360)~\text{keV}$ and 
$Q_\text{EC}=12760(360)~\text{keV}$ in agreement with the $Q_\text{EC}$ values based on the half-life \cite{Fae02} and on the 
systematics \cite{Aud03} (see Table~\ref{tab:2}). The values from the Coulomb displacement energy formula for ${T~=~1}$ in 
\cite{Ant97}, $\Delta E_\text{C}=13552(3)~\text{keV}$ and $Q_\text{EC}=12770(3)~\text{keV}$, agree perfectly with the new fit. 

From the $Q_\text{EC}$ value and the mass of $^{94}$Pd, a mass excess of $-53330(360)~\text{keV}$ is obtained for $^{94}$Ag. 
From the masses of $^{94}$Ag and $^{92}$Rh, a two-proton separation energy in $^{94}$Ag is calculated, $S_\text{2p}= -\Delta 
(^{94}\text{Ag})+ \Delta (^{92}\text{Rh})+ 2\Delta (^1\text{H})=4910(360)~\text{keV}$. If this two-proton separation energy is 
combined with the two-proton-decay data of the high-spin isomer in $^{94}$Ag [$E_\text{2p}=1900(100)~\text{keV}$ \cite{Muk06} 
and $E_\text{x}(^{92}\text{Rh})=1548.6(14)~\text{keV}$ \cite{Kas97}], an excitation energy $E_\text{x}=8360(370)~\text{keV}$ 
is obtained for this isomeric state.  When combined with the measured $^{92}$Rh mass, the two-proton-decay $Q$ value 
\cite{Muk06} gives a mass excess $\Delta=-44970(100)~\text{keV}$ for the high-spin isomer. The newly derived values differ 
significantly from the values based on the systematics [$E_\text{x}=6670(640)\text{\#~keV}$ \cite{Aud03} and 
$\Delta=-46800(500)\text{\# keV}$ \cite{Aud03}] and the empirical shell model [$E_\text{x}=6300~\text{keV}$ \cite{Ple04}]. 
However, the excitation energy is within the extrapolation of $E_\text{x}=6500(2000)\text{\#~keV}$ in Ref.~\cite{nubase}.

\begin{figure}[!]
\center{
\resizebox{0.4\textwidth}{!}{
\includegraphics{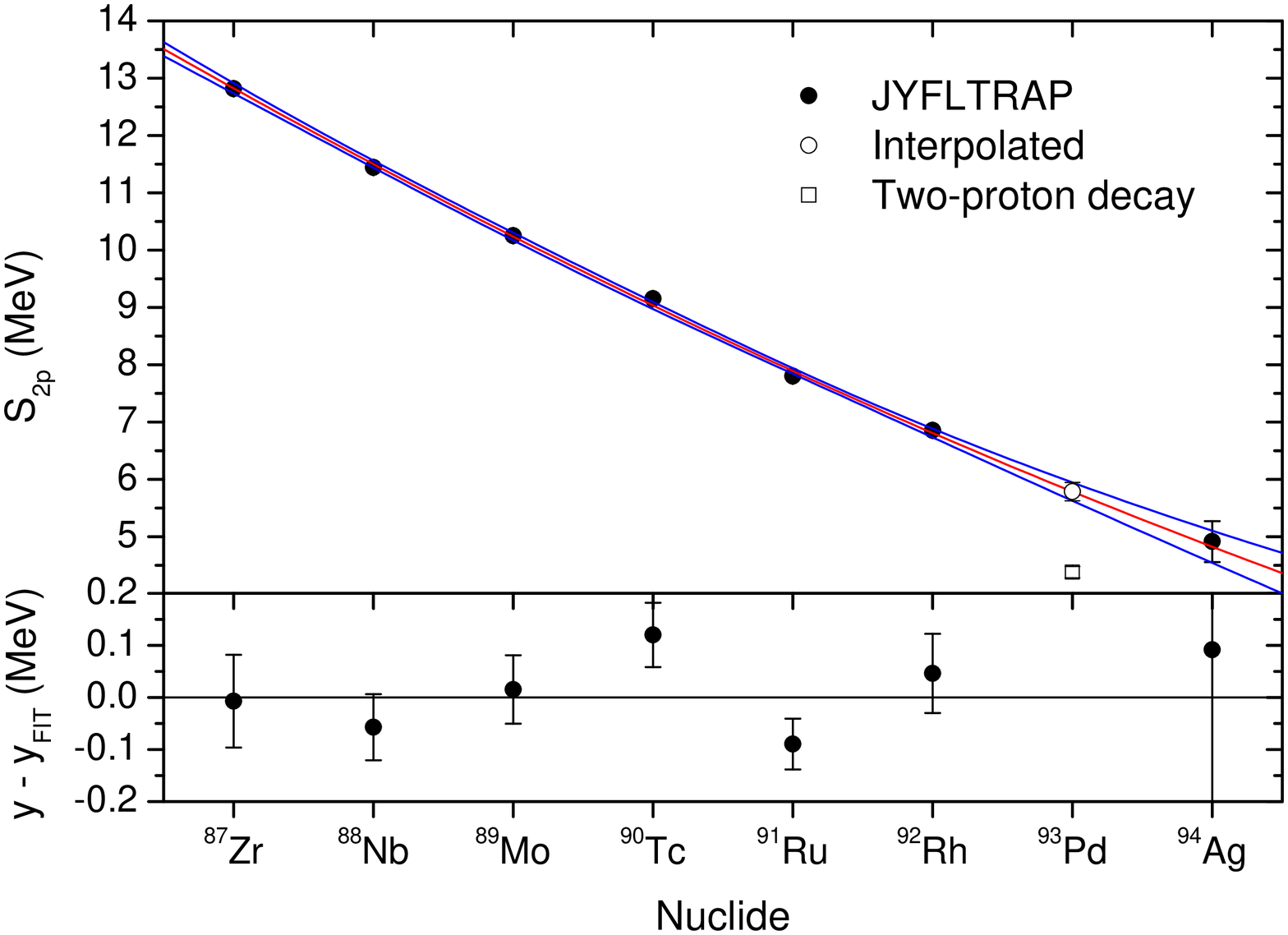}}
\caption{(color online). Two-proton separation energies for the $N=47$ isotones measured at JYFLTRAP. A parabolic fit shown as 
the red line gives $S_\text{2p}=5780(160)~\text{keV}$ for $^{93}$Pd. The $S_\text{2p}(^{93}\text{Pd})= 
-\Delta(^{93}\text{Pd})+\Delta(^{91}\text{Ru})+2\Delta(^1\text{H})=4380(100)~\text{keV}$ using the 
$\Delta(^{93}\text{Pd})=-S_\text{p}(^{93}\text{Pd})+\Delta(^{92}\text{Rh})+\Delta (^1\text{H})$ with 
$S_\text{p}(^{93}\text{Pd})=2330(100)~\text{keV}$ \cite{Muk06} and the measured masses of $^{91}$Ru and $^{92}$Rh shown as an 
open square clearly deviates from the trend. The residuals of the fit are shown in the lower panel.}
\label{fig:3}       
}
\end{figure}

The difference in the isomeric energy between this work and \cite{Muk05} could be due to the extrapolated proton separation 
energy of $^{94}$Ag [$S_\text{p}=-\Delta(^{94}\text{Ag})+\Delta(^{93}\text{Pd})+\Delta(^1\text{H})=890(640)\text{\#~keV}$ 
\cite{Aud03}] which was used together with the proton-decay data 
[$Q_\text{p}=\Delta(^{94}\text{Ag}(21^+))-\Delta(^{93}\text{Pd})-\Delta(^1\text{H})=5780(30)~\text{keV}$ \cite{Muk05}] to 
determine the excitation energy. In order to study this, the mass of $^{93}$Pd was derived from an interpolation of the 
two-proton separation energies in the $N~=~47$ isotones shown in Fig.~\ref{fig:3}. Here the masses from zirconium to rhodium 
were determined at JYFLTRAP \cite{Kan06, Elo08} and the values for $^{85}$Sr and $^{86}$Y are from Ref.~\cite{Aud03}. A 
parabolic fit yields $S_\text{2p}(^{93}\text{Pd})=5780(160)~\text{keV}$, which results in a mass excess of 
$\Delta(^{93}\text{Pd})=-S_\text{2p}(^{93}\text{Pd})+\Delta(^{91}\text{Ru})+2\Delta(^1\text{H})=-59440(160)~\text{keV}$ and 
$S_\text{p}(^{93}\text{Pd})=-\Delta(^{93}\text{Pd})+\Delta(^{92}\text{Rh})+\Delta (^1\text{H})=3730(160)~\text{keV}$. 

\begin{figure}[!]
\center{
\resizebox{0.4\textwidth}{!}{
\includegraphics{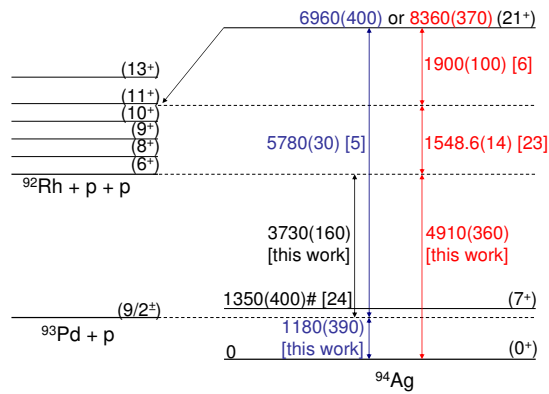}}
\caption{(color online). Decay scheme of $^{94}$Ag ($21^+$). The excitation energy of $^{94}$Ag ($21^+$) depends on whether it 
is determined from the two-proton separation energy of $^{94}$Ag and the two-proton-decay data \cite{Muk06} (shown in red) or 
from the proton separation energy of $^{94}$Ag and the proton-decay $Q$ value \cite{Muk05} (shown in blue). The energies are 
given in keV.}
\label{fig:scheme}       
}
\end{figure}

The mass of $^{93}$Pd yields a proton separation energy of $1180(390)~\text{keV}$ for $^{94}$Ag. If we now combine the 
one-proton decay data \cite{Muk05} with this proton separation energy and the mass of $^{93}$Pd, an excitation energy of 
$6960(400)~\text{keV}$ and a mass excess of $-46370(170)~\text{keV}$ is obtained for the ($21^+$) isomer in agreement with the 
excitation energy given in Ref.~\cite{Muk05}.  Thus, the uncertainty in the extrapolated proton separation energy in 
Ref.~\cite{Aud03} does not explain the observed difference in the excitation energy of the high-spin isomer. The 
two-proton-decay data lead to a significantly different excitation energy than the one-proton-decay data as shown in 
Fig.~\ref{fig:scheme}. A summary and the results for $^{94}$Ag and $^{93}$Pd are given in Table~\ref{tab:2} and compared with 
the values from the AME 2003 \cite{Aud03}.

In order to explain the discrepancy in the excitation energy of $^{94}$Ag ($21^+$), either the one-proton-decay $Q$ value is 
too low or the two-proton-decay energy is too high. In a recent paper \cite{Pec07}, it was claimed that the two-proton decay 
of $^{94}$Ag ($21^+$) to the 1549-keV state in $^{92}$Rh would be highly unlikely, since the most probable state to be 
populated has an excitation energy of $240(570)~\text{keV}$ \cite{Pec07}. With the revised masses of $^{92}$Rh, $^{93}$Pd, and 
$^{94}$Ag ($21^+$) determined in this work, the most probable level to be fed lies at $150(190)~\text{keV}$ when based on the 
one-proton-decay data and agrees with the 1549-keV level in $^{92}$Rh when based on the two-proton-decay data.

The only way to energetically enable the two-proton decay is if the excitation energy of the ($21^+$) isomer in $^{94}$Ag lies 
at $8360(370)~\text{keV}$ instead of $6960(400)~\text{keV}$. A sizable increase in the proton-decay $Q$ value is required to 
explain the $1400(540)~\text{keV}$ difference between the excitation energies. This could be possible if the proton decay fed 
higher-lying states than the 4994-keV and 4751-keV states observed in Ref.~\cite{Muk05}. For example, the 653-keV $\gamma$ 
rays following the de-excitations of the 5648-keV ($37/2^+$) and 6994-keV ($39/2^+$) states \cite{Rus04} may have been hidden 
by a large background of $\beta$-delayed $\gamma$ rays in Ref.~\cite{Muk05}. Another possibility is that some transitions have 
been missed in the decay scheme of $^{93}$Pd in earlier experiments.

\begin{table}[!]
\caption{\label{tab:2}Deduced results for $^{94}$Ag and $^{93}$Pd in comparison with the AME 2003 values based on systematics 
(\#) \cite{Aud03}.}      
\begin{ruledtabular}
\begin{tabular}{rrrrr}
	& 	& JYFL (keV) & AME (keV) & JYFL$-$AME (keV)\\
\hline	
$^{94}$Ag 	& $\Delta$ 		&$-$53330(360) & $-$53300(500)\# & $-$30(620)\\			
			& $Q_\text{EC}$ 	& 12760(360)& 13050(640)\#& $-$290(740)\\
	 		& $S_\text{p}$ 	& 1180(390) & 890(640)\# 	& 290(760)\\
 			& $S_\text{2p}$ 	& 4910(360) & 4520(640)\# & 400(740)\\
\hline			
$^{93}$Pd 	& $\Delta$ 		& $-$59440(160)& $-$59700(400)\# & 260(440)\\
	 		& $S_\text{p}$ 	& 3730(160)& 3630(570)\#& 100(590)\\
			& $S_\text{2p}$ 	& 5780(160)& 5620(710)\# & 160(730)\\
\end{tabular}
\end{ruledtabular}
\end{table}

\begin{table*}[!]
\caption{\label{tab:3} Results for the high-spin ($21^+$) isomer in $^{94}$Ag (given in keV) in comparison with the values 
from the AME 2003 \cite{Aud03} based on systematics (\#). When combined with the data obtained in this work, one-proton- 
\cite{Muk05} and two-proton-decay \cite{Muk06} data lead to different results.}
\begin{ruledtabular}      
\begin{tabular}{rrrr}
$^{94}$Ag ($21^+$)	& One-proton \cite{Muk05} & Two-proton \cite{Muk06} & AME 2003 \\
\hline
$\Delta$ 				& $-$46370(170)	& $-$44970(100)	&  $-$46800(500)\# \\
$E_\text{x}$  			& 6960(400)		& 8360(370)		&  6500(2000)\# \cite{nubase}\\
$Q_\text{EC}$ 			& 19720(170)	& 21130(100)	&  19550(640)\#\\
$Q_\text{p}$ 			& \emph{5780(30)} \cite{Muk05}		& 7180(190)		&  5610(640)\#\\
$Q_\text{2p}$ 			& 2050(170)		& \emph{3450(100)} \cite{Muk06}		& 1980(640)\#\\
\end{tabular}
\end{ruledtabular}
\end{table*}

As the $T~=~1$ state is not always the lowest state in a $T_Z~=~0$ nucleus (for example, in $^{58}$Cu), it is worthwhile to 
consider whether the mass calculated from the Coulomb displacement energies for $^{94}$Ag is a ground-state mass. For example, 
a ($7^+$) isomeric state is expected at $661$ \cite{La02} or $1350(400)\#~\text{keV}$ \cite{nubase} in $^{94}$Ag. If the 
ground state of $^{94}$Ag were the $7^+$ state, this would result in a lower ground-state mass, a higher two-proton separation 
energy, and, hence, an even higher excitation energy of the high-spin ($21^+$) isomer. In contradiction to this latter 
consideration, the ground state is considered to be a $T~=~1$ state \cite{Jan05}.

The excitation energy of $8360(370)~\text{keV}$ for the high-spin isomer lies above the suggested $T~=~1$, ($20^+$) isobaric 
analogue state (IAS) which has been experimentally observed at $7.7~\text{MeV}$ in $^{94}$Pd \cite{Ple04} and should lie at 
the same excitation energy in $^{94}$Ag. Although the higher excitation energy enables a two-proton decay, it is difficult to 
explain why this ($21^+$) state is isomeric and does not rapidly decay to the ($20^+$) state. One explanation is that, due to 
unobserved or wrongly assigned $\beta$-delayed $\gamma$ rays from the $^{94}$Ag ($21^+$) decay, the observed ($20^+$) level in 
$^{94}$Pd may in reality be a $18^+$ state which would mean that the ($20^+$), $T=1$ state would lie higher in both $^{94}$Pd 
and $^{94}$Ag. The high-spin isomer has been suggested to be highly deformed in Ref.~\cite{Muk06} but this has later been 
questioned by the calculations in Refs.~\cite{Pec07, Kan08}. If the ($20^+$) IAS lies below the ($21^+$) high-spin isomer and 
a possible high deformation of this isomer does not explain the hindrance of the internal decay, the excitation energy of the 
isomer has to be lower, supporting the one-proton-decay data. 
 
In conclusion, one-proton-decay data \cite{Muk05} and two-proton-decay data \cite{Muk06} disagree with each other when the 
mass excess values for $^{92}$Rh, $^{93}$Pd and $^{94}$Ag are combined with these data. A possible explanation for this 
discrepancy is that some $\gamma$ transitions have not been observed in the one-proton decay due to a large background of 
$\beta$-delayed $\gamma$ rays. This would suggest that the excitation energy of the $^{94}$Ag ($21^+$) isomer lies at around 
$8.4~\text{MeV}$ instead of $7.0~\text{MeV}$. Although the excitation energy of the high-spin isomer in $^{94}$Ag remains 
uncertain, we have obtained new data on its decay $Q$ values tabulated for the two different excitation energies obtained in 
this work in Table~\ref{tab:3}. In addition, we have considerably reduced the uncertainties of the one-, two- and three-proton 
separation energies in $^{94}$Pd needed in the decay studies of $^{94}$Ag ($21^+$). Further experiments on the one-proton 
decay of $^{94}$Ag  ($21^+$) and on the decay schemes of $^{92}$Rh and $^{93}$Pd could verify the proton-decay $Q$ value. To 
finally solve this puzzle, direct mass measurements on $^{93}$Pd, $^{94}$Ag and $^{94}$Ag ($21^+$) are needed, posing new 
challenges for the production of these exotic species. These challenges are presently being pursued at the IGISOL facility 
\cite{Kes07}.

\begin{acknowledgments}
This work has been supported by the EU 6th Framework program ``Integrating Infrastructure Initiative - Transnational Access", 
Contract Number: 506065 (EURONS) and by the Academy of Finland under the Finnish Center of Excellence Program 2006-2011 
(Nuclear and Accelerator Based Physics Program at JYFL). The support via the Finnish-Russian Interacademy Agreement (project 
No. 8) is acknowledged. The authors thank Ernst Roeckl for his valuable comments.
\end{acknowledgments}

\bibliography{aps}
\end{document}